\documentclass[superscriptaddress,bibnotes,twocolumn,aps,prb]{revtex4}
\usepackage{graphicx}
\usepackage{bm}
\usepackage{float}
\usepackage{dcolumn}

\begin{document}

\title{Bistable hysteresis and resistance switching in hydrogen-gold junctions}
\author{M.L. Trouwborst}
\altaffiliation{Present address: IBM Research - Z\"{u}rich. Email address: mtr@zurich.ibm.com}
\affiliation{Physics of Nanodevices, Zernike Institute for Advanced
Materials, Rijksuniversiteit Groningen, Nijenborgh 4, 9747 AG
Groningen, The Netherlands}
\author{E.H. Huisman}
\affiliation{Physics of Nanodevices, Zernike Institute for Advanced
Materials, Rijksuniversiteit Groningen, Nijenborgh 4, 9747 AG
Groningen, The Netherlands}
\author{S.J. van der Molen}
\affiliation{Kamerlingh Onnes Laboratorium,  Leiden University, P.O.
Box 9504, 2300 RA, Leiden, The Netherlands}
\author{B.J. van Wees}
\affiliation{Physics of Nanodevices, Zernike Institute for Advanced
Materials, Rijksuniversiteit Groningen, Nijenborgh 4, 9747 AG
Groningen, The Netherlands}
\date{\today}
\begin{abstract}
Current-voltage characteristics of $H_2-Au$ molecular junctions
exhibit intriguing steps around a characteristic voltage of
$V_s\approx40$ mV. Surprisingly, we find that a hysteresis is
connected to these steps with a typical time scale $> 10$ ms. This
time constant scales linearly with the power dissipated in the
junction beyond an off-set power $P_s=IV_s$. We propose that the
hysteresis is related to vibrational heating of both the molecule
in the junction and a set of surrounding hydrogen molecules.
Remarkably, by stretching the junction the hysteresis' characteristic time becomes $>$ days. We demonstrate
that reliable switchable devices can be built from such junctions.
\end{abstract}
\maketitle
The central idea behind molecular electronics is that novel
devices can be created using intelligently designed functional
molecules. The simplest example is a molecular junction that
exhibits conductance switching. Chemists have synthesized various
types of bistable molecules, many of which have subsequently been
applied in devices \cite{feringa,dulic,he,meyer,molen}.
Surprisingly, however, conductance switching has also been
observed in junctions that incorporate passive molecules only
\cite{weiss,blum,emanuel,lathaNN}. Understanding the fascinating
physics behind the latter phenomenon may lead to a novel design
philosophy of switchable molecular devices. Here, we present
detailed transport studies on $H_2-Au$ molecular junctions. We
demonstrate that if I(V)-curves are taken by fast data
acquisition, hysteretic behavior is observed. Moreover, we can
manipulate our junctions such that the hysteresis' characteristic
time scale grows from $\sim$10 ms to $>$days. The resulting
$H_2-Au$ junctions are arguably the smallest switchable devices
around. The physics behind these fascinating phenomena forms the
main subject of this Letter.

\begin{figure}[!h]
\begin{center}
\includegraphics[width=8.5cm]{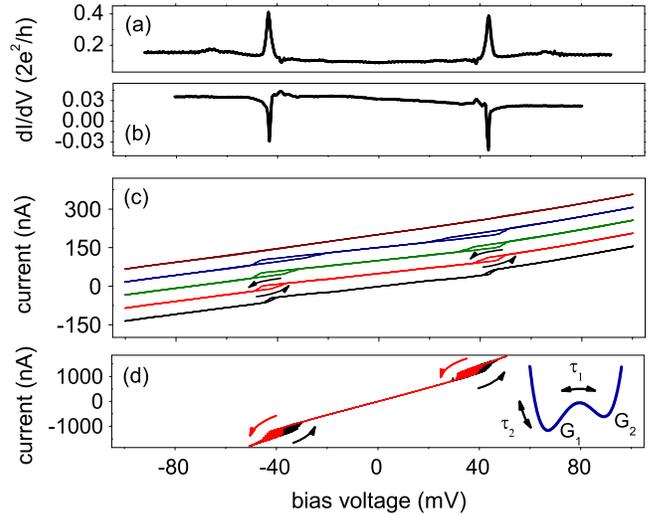}
\end{center}
\caption{\textit{a),b) Examples showing peaks (a) or dips (b) in
dI/dV around $\approx 43$ mV, measured on a $H_2-Au$ contact (at 5 K). Note that (b) features a negative differential
conductance. c) I(V)-curves measured with varying sweep rates. The lower curve, measured
in 10 s, corresponds to curves such as a). However, for faster
voltage sweeps, a hysteresis shows up around $V_s$. Sweeping times
are 10, 1, 0.5, 0.33 and 0.1 s. For
clarity, the traces are shifted vertically. d)
I(V)-curve showing both hysteresis and two level fluctuations
around $V_s$ (voltage swept in 1s). Inset: double
well potential for the two-level model. We associate the fast switching between the two wells to time scale
$\tau_1$, and the slower changes in the potential landscape to $\tau_2$.}} \label{figure1}
\end{figure}

\begin{figure}[h]
\begin{center}
\includegraphics[width=8.5cm]{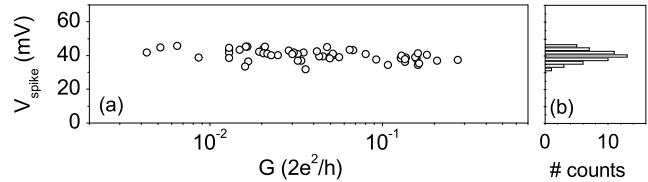}
\end{center}
\caption{\textit{a) Position of the conductance peak/dip for 56
different contacts as a function of its zero bias conductance.
b) Histogram of all data points. On average, $V_s=40\pm 3$ mV.} }
\label{histogram}
\end{figure}

Molecular junctions incorporating hydrogen molecules have a number
of remarkable properties
\cite{Nature_Roel,gupta,thijssenprl,halbritter,temirov}. Perhaps
most striking is that differential conductance (dI/dV) curves
exhibit sharp features at a well-defined voltage $V_s$
\cite{gupta,thijssenprl,halbritter,temirov}. In Fig.\ref{figure1},
we show two typical measurements for a $H_2-Au$ contact. In one
case, two sharp peaks are observed (Fig.\ref{figure1}a), whereas
in the other, two dips are seen (Fig.\ref{figure1}b). Both curves
are symmetric with respect to voltage. Remarkably, the dip in
Fig.\ref{figure1}b) is so deep that the differential conductance
around $V_s=\pm$43 mV becomes negative. Recently, Thijssen and
Halbritter \textit{et al.} measured curves similar to Fig.
\ref{figure1}a) and b) and put forward a two-level model
\cite{thijssenprl,halbritter}. The two levels (1, 2), most likely
representing different geometries of the $H_2-Au$ contact, give
rise to two specific conductance values ($G_1$, $G_2$)
\cite{barnett}. Hence, if transitions between level 1 and level 2
are induced at $V=V_s$, the conductance changes discontinuously
from $G_1$ to an average of $G_1$ and $G_2$ (the weighing being
dependent on the degeneracy of the levels \cite{halbritter}). This
naturally results in a dip (if $G_1>G_2$) or peak (if $G_1<G_2$)
in dI/dV, or analogously, in a step down or up in an I(V)-curve.
The latter is illustrated by the lowest I(V)-characteristic
presented in Fig. \ref{figure1}c). In both models proposed, a
double potential well is assumed to describe the two-level system
(schematized in Fig.\ref{figure1}d), inset). However, there is
disagreement on the question how a transition between the levels
occurs. Halbritter \textit{et al.} relate $V_s$ to the energy
difference between the two levels. Furthermore, they explain the
negative differential conductance by proposing that level 2 is
highly degenerate. In contrast, Thijssen \textit{et al.} propose
that a transition is only possible if a molecular vibration is
excited with high enough energy $\hbar \omega_1$ to assist a jump
over the barrier. Hence, they conclude that $V_s=\hbar
\omega_1/e$, a statement for which they provide significant
experimental evidence. Regardless of the exact model, however, the
transition between states 1 and 2 is anticipated to be fast around
$V_s$, with time scales $<1$ ns, determined by tunnelling or
electron-phonon interaction. The results we present here, strongly
contradict this expectation. In Fig.\ref{figure1}c), we show
I(V)-curves taken with various sweeping times.  When slowly
ramping V, one obtains the lower curve discussed above. However,
when sweeping faster than 1 s per curve, a pronounced hysteresis
appears around $V=V_s$. In other words, there is a macroscopic
time scale connected to the system's return from the high
conductance state to the low one. Moreover, for even faster
voltage sweeps, $\geq 10$ Hz, the system does not have enough time
to relax back to state 1; it keeps the 'high' conductance value
for all V. Not only does this not agree with the two-level models,
it also rules out the possibility that the steps in the I(V)'s are
due to resonant tunnelling. Hence, this intriguing
hysteresis forms the central item of this study.\\
To obtain stable molecular junctions, we employ mechanically
controllable break junctions \cite{agrait}. Basically, a gold wire
is broken in cryogenic vacuum (at $\approx$5 K), and the electrode
tips are sharpened by a "training" procedure \cite{PRL_G0loop}.
Subsequently, a small volume of $H_2$ (99.999 $\%$ pure) is
introduced and the junction is opened with $\approx$5 pm/s
\cite{vrouwe}. Meanwhile, I(V)-curves are recorded by sweeping V
between $-90$ and $90$ mV at $\approx 1$ Hz. After several
minutes, nonlinear I(V)-curves are observed with pronounced
conductance steps \cite{thijssenprl,halbritter}. During this
procedure, we do not observe a jump out of contact. In total, we
have used 6 break junctions on which 56 different $H_2-Au$
contacts were studied. To obtain geometrically different $H_2-Au$
junctions for one single device, we use the following procedure.
First, the $H_2-Au$ junction is broken, resulting in a jump out of
contact. Then, the electrodes are pushed back into contact to
conductance values $>10 G_0$ (where $G_0=2e^2/h$). Finally, the
gentle pulling process is started again until a new $H_2-Au$
contact forms. To connect to previous work, we have first
investigated the I(V) curves of all these junctions at low voltage
scan rates. In Fig. \ref{histogram}a) we present a statistical
analysis of the on-set voltage $V_s$. Interestingly, $V_s$ is
positioned around 40 meV, rather independent of the conductance of
the contact. Indeed, the zero bias conductances show a broad
variation between 0.003 and 0.3 $G_0$. Since we did not observe a
jump out of contact while creating the junctions, this may point
to chain formation as reported in Ref. \cite{csonka5}. These authors find
conductance values in the same range as observed here. In Fig.
\ref{histogram}b), we present a histogram of all $V_s$ values. The
average position is $V_s=40\pm 3$ mV. This is in good agreement
with Thijssen \textit{et al}, who found $V_s=42 \pm 2$ mV
\cite{thijssenprl}. As explained above, they relate $V_s$ to a
$H_2$ phonon, which induces transitions between levels 1 and 2.
Experimentally, they provide two pieces of evidence.
First, they show that $V_s$ shifts down by a factor $\approx
\sqrt{2}$ when $H_2$ is replaced by $D_2$ (deuterium). Second,
they show that upon pulling a junction, a step down in a
dI/dV-curve can evolve into a peak. Since a step in dI/dV is
related to the excitation of a phonon \cite{oren,Thygesen}, this confirms
their assertion. We have performed similar pulling experiments. In
some cases, peaks in dI/dV grew larger. In other cases, they would
virtually disappear, leaving a step as the dominant feature. This
confirms the result by Thijssen.

As we increase the scan velocity of our I(V)-scans, a hysteresis
develops around $V_s$ (Fig.\ref{figure1}c). Generally, we observe
a smooth transition from one conductance state to the other.
However, this is not always the case. The I(V)-scan in
Fig.\ref{figure1}d) for example, shows pronounced two levels
fluctuations (TLFs) in combination with hysteresis. A comparison
of Figs.\ref{figure1}c) and d) leads to a number of important
observations. First, the hysteresis is not an artifact of our
electronics: TLFs much faster than the time scale of the
hysteresis are easily observed. Second, the occurrence of TLFs
confirms a two-level model. Halbritter \textit{et al.} could not
resolve the actual TLFs. Also in our case, most curves show an
apparently smooth step around $V_s$. This is due to the TLF
frequency exceeding the bandwidth of the electronics ($\approx 50$ kHz). Third, we see that the TLF-amplitude around $V_S$ is
approximately equal to the difference between the 'low' and 'high'
conductance states far away from $V_s$. This indicates that for $V
\gg V_s$, the system is in state 2 most of the time. Hence, state
2 is either relatively stable or highly degenerate
\cite{halbritter}. Finally, we deduce from Fig. \ref{figure1}d)
that despite continuous communication between both levels (through
the TLFs), the hysteresis is still present. This suggests that the
exact shape of the double potential well (at $V \approx V_s$)
differs for the upward and downward voltage scans. We deduce that
there are two time scales in our system: a time $\tau_1$
corresponding to fast switching between states (TLFs) and $\tau_2
> \tau_1$ connected to changes in the effective potential
landscape, resulting in hysteresis.

\begin{figure}[!h]
\begin{center}
\includegraphics[width=8.5cm]{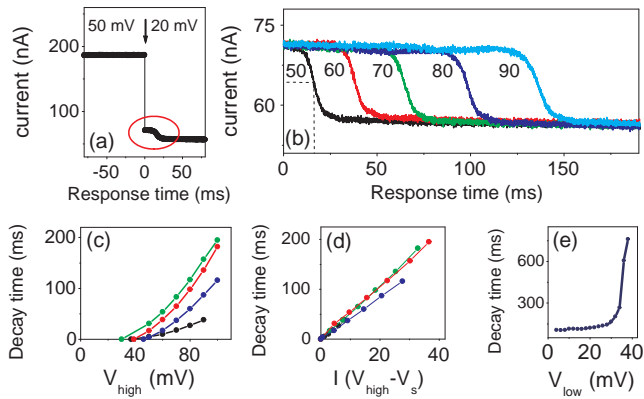}
\end{center}
\caption{a) \textit{Pulse response. At t=0, the voltage is
instantly dropped from a high voltage ($V_{high}>V_{s}$), to a
low voltage ($V_{low}<V_{s}$). Here, $V_{high}$
($V_{low}$) is 50 mV (20 mV). Right after the voltage drop, the
conductance is still in the high conductance state. b)
Response for different initial voltages. $V_{high}$ is
systematically varied between 50 mV (first curve) and 90 mV (last
curve), while $V_{low}$ is 20 mV. The decay time $\tau_2$ is defined as the time where the
conductance has dropped halfway (16 ms for $V_{high}=$50 mV). c) Decay times as a function of $V_{high}$,
for 4 different contacts ($V_{low}=20$ mV.) The
contacts have a conductance of, from lower to upper curve, 0.015,
0.044, 0.048 and 0.058 $G_0$ (at 50 mV). d) Decay times as a
function of dissipated power during the voltage pulse, corrected for an onset
value $P_s=IV_s$ (same contacts as in c). e) Decay times as a function of $V_{low}$.
$V_{high}$ was set to 80 mV and for this contact $V_s=$39 mV.}} \label{pulsen}
\end{figure}

To gain more insight in $\tau_2$, we have recorded response times
using well-defined voltage pulses. An example is given in
Fig.\ref{pulsen}a). First, V is set to a value $V_{high}>V_s$,
i.e. beyond the conductance step. The system's conductance is
'high'. After 1 s, long enough to reach equilibrium, the voltage
is instantly dropped to a value $V_{low}<V_s$ (in this case 20
mV). Interestingly, the current responds to this decrease within
20 $\mu s$, consistent with our bandwidth. However, the
\textit{conductance} $I/V$ does not. It is still 'high' after 20
$\mu s$. Only after 10 ms, it starts to drop to a lower value. We
define $\tau_2$ as the time in which the conductance drops halfway
(dotted line in Fig.\ref{pulsen}b). Remarkably, we find larger
$\tau_2$ for higher $V_{high}$ (see Fig.\ref{pulsen}b), which
implies that $\tau_2$ depends on the history of the junction. We
have systematically investigated $\tau_2$ vs. $V_{high}$ for 4
different contacts, as depicted in Fig.\ref{pulsen}c) \footnote{Only voltage pulses $<$100 mV could be used since higher voltages led to irreversible changes in the molecular contact.}. Clearly, $\tau_2$ increases more than linearly with $V_{high}$, with an
onset around 40 meV. In fact, if we plot $\tau_2$ as a function of
$I \cdot (V_{high}-V_s)$ we find a linear relation for all 4
samples (see Fig. \ref{pulsen}d). Hence, there is a direct
connection between $\tau_2$ and the power $IV_{high}$ that was
being dissipated before the voltage step, corrected for an onset
value $P_s=IV_s$. From this interesting result, we deduce that
heating effects (multiple phonon excitations) are at the basis of
the hysteresis. The fact that $V_s=40$ mV serves as a threshold in
all cases points to the pivotal role of a hydrogen phonon in this
process \cite{thijssenprl}. A second experiment that demonstrates
this connection is shown in Fig. \ref{pulsen}e). This measurement
is similar to that in Fig. \ref{pulsen}b). However, now $V_{high}$
is kept constant at 80 mV, whereas $V_{low}$ is systematically
varied. We find that $\tau_2$ increases dramatically as $V_{low}$
approaches $V_s$.

\begin{figure}[!h]
\begin{center}
\includegraphics[width=8.5cm]{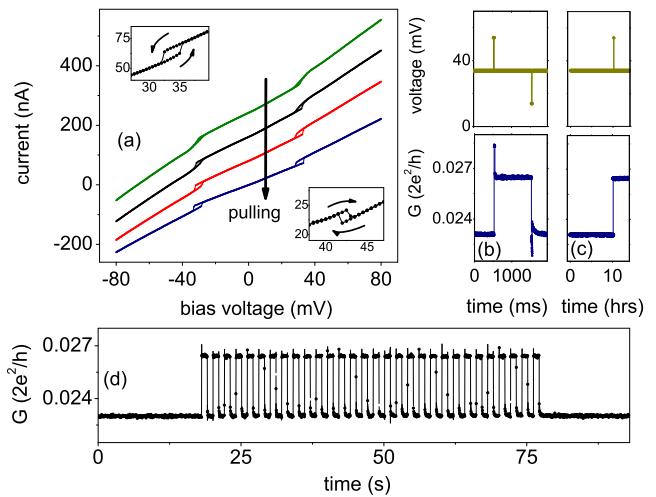}
\end{center}
\caption{\textit{a) I(V) sweeps while pulling the electrodes
(voltage is swept in 1 min). After stretching $2 \pm 0.3 \AA$, a
time independent hysteresis appears. For clarity, the curves are
shifted vertically. Insets: two other examples of contacts showing
time independent hysteresis. b) Hydrogen switch. Upper graph:
Voltage used to control the switch. On top of an offset voltage
(here 34 mV), the voltage is pulsed with +20 mV and -20 mV
(duration is 20 ms). Lower graph: Pulse response. After each
pulse, the system switches. c) Stability of the hydrogen switch.
After 10 hours, the voltage is pulsed by +20 mV (20 ms) to switch
to $G_2$. After 14 hours, the measurement is stopped. d) Multiple
switching events. The pulse height is $\pm$4 mV, on top of a dc
voltage of 34 mV (pulse duration is 20 ms).} } \label{switch}
\end{figure}

Remarkably, when slowly stretching the contact, a special type of
hysteresis appears for the majority of the junctions. In general, a contact is stretched in steps of 10 pm while for each position an I(V) curve is measured (in 1 min). Fig.\ref{switch}a) shows 4 I(V)'s with 50 pm intervals. The upper
curve is measured right after the $H_2-Au$ contact is formed.
With further stretching a hysteresis appears, which is most
pronounced after 2 $\AA$ (lower curve). Interestingly, this type of hysteresis does not show any time
dependence, contrasting the data above. Therefore, we can use it
as a stable memory element. For this, we first apply an offset
voltage $V=V_s$. Subsequently, we switch conductance by applying
positive (low G $\rightarrow$ high G) or negative voltage pulses
(v.v.). This is demonstrated in Fig.\ref{switch}b. The switch is
remarkably stable in time. Even after 14 hours, no degradation is
observed (see Fig. \ref{switch}c). Furthermore, no deterioration
is seen after multiple switching events (see Fig. \ref{switch}d).
We have performed over 5000 switching events, and saw no changes
to the sample. Hence, we have created a switchable molecular
device from the smallest molecule nature provides.

To discuss our observations, we first return to Fig.\ref{pulsen}.
Here we concluded that heating, i.e. electron-phonon excitation,
plays a pivotal role in the hysteresis. Furthermore, the molecular
phonons are crucial in this process. We never observed heating
effects on bare gold electrodes, which is consistent with the
relatively low dissipation in our junctions (cf. Ref.
\cite{kolesnychenko}). Recently, Galperin \textit{et al.} and Chen
theoretically studied the temperature of molecular junctions under
a bias voltage \cite{galperin3,chenn}. It is not trivial, however,
to define a temperature for a single molecule. Galperin \textit{et
al.} approach this problem as follows. First, they assume that an
external phonon bath is weakly coupled to the active mode of the
molecule in the junction. Then, they calculate the ensuing heat
flux between mode and bath. The bath temperature at which the heat
flux vanishes is finally assigned to the molecular junction. For
$H_2$-junctions, a temperature increase $\sim 100$ K was
predicted, exactly when the voltage reaches the phonon energy
$eV=\hbar\omega$ \cite{galperin3,chenn}. From this, one might
naively expect a hysteresis between upward and downward
I(V)-scans. Going up in V, the molecule is 'cold' and in state 1;
coming down, the molecule is 'hot', i.e. in a combination of
states 1 and 2. However, it is highly unlikely that a single
molecule will remain excited on time scales as large as
$\tau_2>10$ ms. For the most likely relaxation mechanism, electron
phonon coupling, typical time scales are $<1$ ns, depending on the
vibration mode \cite{pecchia}. We propose that there is a larger
phonon bath that preserves the energy for a relatively long time
(cf. Galperin). A good candidate is formed by the large set of
$H_2$ molecules that likely surround the actual molecule in the
junction. At $V>V_s$, these molecules will be excited by efficient
phonon-phonon coupling from the junction outwards. After V is
dropped below $V_s$, it will take a relatively long time before
this bath (and the molecule in the junction) is at the base
temperature again. It may well be that heating of surrounding
$H_2$ molecules actually induces a phase transition, although this
is not essential to explain Figs. \ref{figure1}-\ref{pulsen}. In fact, we have only observed current steps at experimental
temperatures below $\approx20$ K. A phase transition has been proposed by Gupta and Temirov \textit{et al.}
to explain non-linear IV curves \cite{gupta,temirov}. Such a
transition could indeed explain the time-independent hysteresis
observed in Fig. \ref{switch}. Upon a phase transition, the
hydrogen phonon energy will most likely change. If it assumes a
lower value $\omega_2 $ than before ($\omega_1$), the hydrogen
molecule(s) will still be excited if the voltage drops below
$V_s$, provided that $\omega_2<eV<\omega_1$. This situation may be
quasi-permanent, since the system is continuously excited. Only
when V is decreased further, a transition back is expected. Our
work may thus represent the first experimental case in which
molecular heating is directly distinguished in a single junction
\cite{galperin3,chenn}.

In summary, we demonstrate that the presence of $H_2$ in a Au
junction leads to two-level fluctuations, steps in I(V)-curves and
hysteresis. Moreover, by carefully stretching the junctions, a
reversible and stable molecular switch can be created.
Interestingly, similar features have been reported for a large
variety of molecular junctions, with molecules ranging from larger
organic molecules to CO \cite{weiss,blum,emanuel,thijssenprl,gaudioso,danilov}.
Since such switching effects are general phenomena in molecular
electronics, there may be one basic mechanism behind them. In
parallel, our work stresses the requirement to work in a
hydrogen-free environment if one focuses on a different molecular
system. Finally, we note that the steps in I(V)-curves (explained
by a two-level model) may easily be confused with resonant
tunnelling. To distinguish the two, fast I(V)-scanning is an
efficient tool, since resonant tunnelling should not lead to
hysteresis.

This work is financed by the Nederlandse Organisatie voor
Wetenschappelijk onderzoek, NWO (Pionier grant), and the FOM program Atomic and Molecular
nanophysics. We thank Frank Bakker, Simon Vrouwe, Siemon Bakker and Bernard Wolfs for
technical support and Jan van Ruitenbeek and Roel Smit for discussions.

\end{document}